# The Greggs-Pret Index: a Machine Learning analysis of consumer habits as a metric for the socio-economic North-South divide in England

**Robin Smith[1], Kristian C. Z. Haverson[1]**

[1]Department of Engineering and Mathematics, Sheffield Hallam University, Howard Street, Sheffield, S1 1WB, UK
Email: robin.smith@shu.ac.uk

April 1, 2023

## ABSTRACT

In England, it is anecdotally remarked that the number of Greggs bakeries to be found in a town is a reliable measure of the area's 'Northern-ness'. Conversely, a commercial competitor to Greggs in the baked goods and sandwiches market, Pret-a-Manger, is reputed to be popular in more 'southern' areas of England. Using a Support Vector Machine and an Artificial Neural Network (ANN) Regression Model, the relative geographical distributions of Greggs and Pret have been utilised for the first time to quantify the North-South divide in England. The calculated dividing lines were each compared to another line, based on Gross Domestic Household Income (GDHI). The lines match remarkably well, and we conclude that this is likely because much of England's wealth is concentrated in London, as are most of England's Pret-a-Manger shops. Further studies were conducted based on the relative geographical distributions of popular supermarkets Morrisons and Waitrose, which are also considered to have a North-South association. This analysis yields different results. For all metrics, the North-South dividing line passes close to the M1 Watford Gap services. As a common British idiom, this location is oft quoted as one point along the English North-South divide, and it is notable that this work agrees. This tongue-in-cheek analysis aims to highlight more serious factors highlighting the North-South divide, such as life expectancy, education, and poverty.

## 1. Introduction

The concept of an English North–South divide pervades English history: a line that dissects the country, rendering the two halves socially, economically, and culturally distinct. Some historians trace this back as far as the year 1069 when, "like a raging lion" [1], William the Conqueror charged northwards in an attempt to control the more unruly parts of his new kingdom [2]. Ever since then, as the common idiom goes, the North of England has suffered more than South: this speculation has inevitably led to a number of stereotypes, which we will not discuss here.

Greggs is a British bakery chain. It specialises in savoury products such as pastry bakes, sausage rolls, sandwiches and sweet items including doughnuts and cookies. Greggs was founded by John Gregg as a small Tyneside bakery in 1939. It opened its first shop in Gosforth, Newcastle upon Tyne in 1951. Owing in part to its geographical origins, it is a brand associated with the North of England. From these humble beginnings, the business has grown and Greggs is now the largest retail baker in the UK with stores across the whole country. Recent rapid growth can in part be attributed to the successful release of its famous vegan sausage roll in early 2019 [3].

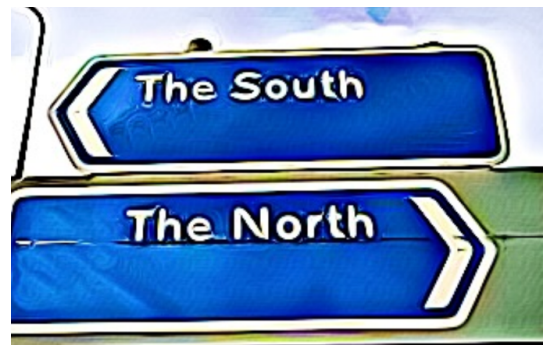

**Figure 1. The North-South dividing line has long been an issue of contention in England.**

In contrast, Pret-a-Manger ('Pret') is a British sandwich shop chain founded by Jeffrey Hyman in 1983. The first shop opened in Hampstead – a London neighbourhood and one of the wealthiest areas of England – in 1984. The company went into liquidation in 1986 and the name and visual branding were purchased from the company liquidator by Sinclair Beecham and Julian Metcalfe, who have grown the business to £461.5m revenue in 2021.

With Greggs and Pret each having geographical origins in the North and South of England, respectively, it is understandable that 'Northern-ness' and 'Southern-ness' are often associated with each of these brands. In 2017, student newspaper The Tab posted a tongue-in-cheek article claiming to have defined the England North-South divide by plotting all English Greggs shops on a map [4]. Their calculations rested on the idea that the fewer people per Greggs shop in a city, the more northern that place is. Their chosen tipping point was cited as 25,000 people per shop. Any town or city above that number was allocated as "southern". Their work contentiously included Scottish and Welsh cities into their calculations, which our work excludes from the English North-South debate.

This article builds on the work of [4] by attempting to define the English North-South divide by the relative geographical distributions of both Greggs and Pret shops. This was achieved using supervised Machine Learning algorithms including a Support Vector Machine and an Artificial Neural Network (ANN) Regression Model via the Matlab Statistics and Machine Learning Toolbox. These methods are described in more detail in the following section. The results were compared with an alternative dividing line, derived by an ANN and trained on Gross Domestic Household Income (GDHI) of the areas of England at International Territorial Levels, Level 2 [5]. An alternative definition of the North-South divide based on the relative geographical distributions of Morrisons and Waitrose stores is also included.

## 2. Methodology

### 2.1. Support Vector Machine

A Support Vector Machine (SVM) is a supervised computer algorithm that learns by example to assign labels to objects [6,7]. These SVMs have been successfully applied to an increasingly wide variety of applications across research and within society. They were most famous shortly after their invention in the 1990s, but continue to be the go-to method for a high-accuracy classification algorithm.

The algorithm takes a set of input data features. In our case this is the latitude and longitude coordinates of every Greggs and Pret store in England. Each data point is labelled with a category; in our case this is "Greggs, Northern $(+1)$" and "Pret, Southern $(-1)$". The aim of the SVM is to find a hyperplane that maximally separates the different classes in the training data. For our two-dimensional latitude and longitude inputs, this is just a line. The algorithm finds an optimal hyperplane that separates the data by maximising the margin, which is defined as the distance between the hyperplane and the closest data points from each class. One difficulty arises because there are many more Greggs shops than Pret, and the SVM is designed to operate with equal numbers of data in each category. To overcome this, a random sample set of Greggs data were repeatedly taken to match the sample size of Pret stores. The optimal hyperplane was then determined for each sample set. This was performed with 1000 sample sets and an average hyperplane was calculated to define the North-South divide.

The SVMs are particularly useful when the data to be classified has many features or when there is a clear margin of separation in the data. The result for the Greggs-Pret data is a straight line that passes through England that best separates the geographical locations of the Greggs and Pret Stores. This work utilises the standard Matlab SVM model [8].

### 2.2. Neural Network Regression Model

Artificial Neural Network Regression Models [9] predict an output variable as a function of a set of inputs. The input features (independent variables) of a classification neural network can be categorical but for Regression Models the inputs must be numeric. For this application, there are just two input variables: the latitude and longitude coordinates of every Greggs and Pret shop in England. The output was chosen to be 1 for Pret and 0 for Greggs locations. The Artificial Neural Network used was the standard Matlab model [10].

Neural Networks consist of a set of connected layers, each with several nodes. The set of input variables (coordinates) are combined via mathematical formulae as they propagate through the neural network from layer to layer and are multiplied by certain weights and added to certain offsets. During the training stage these weights and offsets are varied such that the difference between the outputs of the network and the outputs of the training data is minimised. Again a difficulty arises because there are many more Greggs stores than Pret, and the Neural Network functions best with equal numbers of data in each category. To overcome this, a random sample of Greggs data were repeatedly taken to match the sample size of Pret stores. Then for each sample, to determine England's 'Northern-ness' or 'Southern-ness', a grid of uniformly separated coordinates throughout England are provided as an input to the trained Neural Network. For each pair of coordinates, a value is calculated. A value close to 1 indicates the South and a value close to 0 represents the North. This process was repeated with 1000 random samples and an average map was calculated. The locations on the map with a value of 0.5 were used to define a contour line separating North from South.

This method was also applied to GDHI data. Here the latitude and longitude coordinates of all areas of England at International Territorial Levels, Level 2 (ITL2) [5] were selected as the inputs to the Neural Network and the output was chosen to be the mean GDHI for that region. The GDHI values were linearly scaled such that the minimum GDHI was zero and the maximum GDHI was 1. Again, a grid of uniformly separated coordinates throughout England was then provided as an input to the trained Neural Network. For each pair of coordinates, a value was calculated. A value close to 1 indicates a predicted high wealth and a value close to 0 represents low wealth. The mean GDHI for England in 2021 (£21,962) was then used to define a contour line, separating North from South.

## 3. Results

### 3.1. Greggs-Pret comparison

Initially, 80% of the data were used for training the models and 20% were kept for testing the models. The trained SVM had an average accuracy of 78%, meaning that when applied to the remaining test data, 78% were correctly allocated into a Greggs (North) and Pret (South) category. The distribution of Greggs and Pret shops, along with the optimal dividing line, are shown in Figure 2.

The ANN performed similarly well, obtaining an accuracy of 81% when applied to the test data. The line predicted for the North-South separation by this model is shown in Figure 3. It is akin to the prediction of the SVM in the linear region. However, it dives southwards towards the end of its trajectory. It assigns the top of Norfolk, including King's Lynn and the Norfolk Coast Area of Outstanding Natural Beauty, as Northern. Norwich avoids this and remains classified as Southern.

It is notable that the predicted line from both of these calculations passes close to the Watford Gap services on the M1 motorway. This has long been touted in British conversation as a location that lies along the English North-South divide.

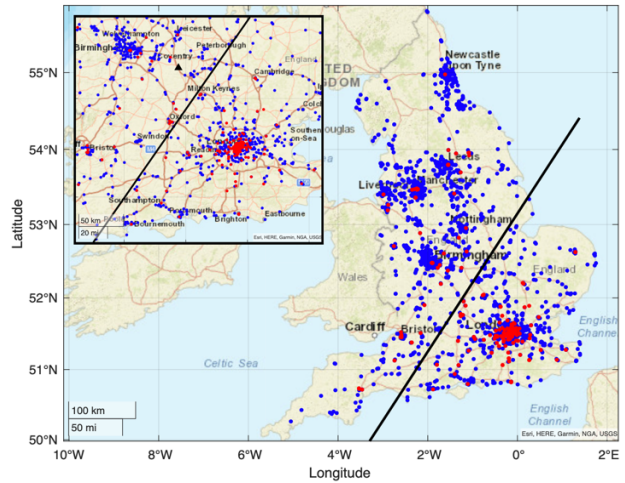

**Figure 2.** The North-South dividing line as determined by the Greggs and Pret shops analysed using the SVM. Greggs shops are plotted in blue, and Pret in red.

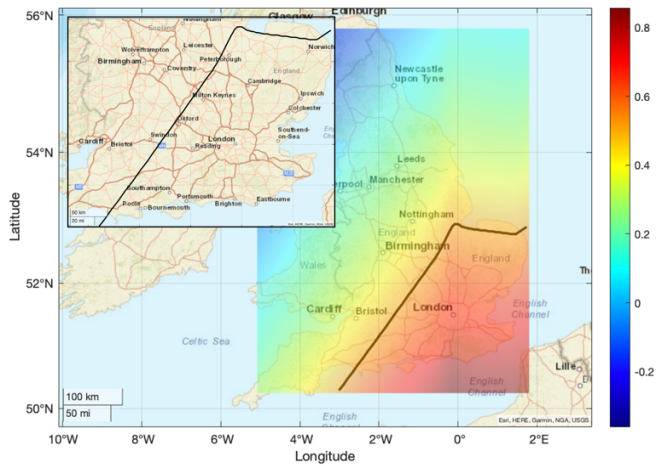

**Figure 3.** The North-South dividing line as determined by the Artificial Neural Network regression analysis of Greggs and Pret locations. The heatmap denotes the calculated 'Northern-ness' and 'Southern-ness' as a function of location. The black line is a separating contour line.

### 3.2. Gross Domestic Household Income

As in the previous analyses, 80% of the GDHI data were used for training the ANN and 20% were kept for testing. The trained ANN had an average accuracy of 95%, meaning that the neural network was able to predict GDHI based on the coordinates of a place in England (within the 20% sample of test data) to 95% accuracy. However, there were only 43 English locations in the International Territorial Levels data set, meaning that statistics are limited.

In the same way as for section 3.1., a grid of latitude and longitude coordinates across the landmass of the UK was generated. The trained ANN was then used to predict the GDHI at these locations. A contour line, corresponding to the mean GDHI for England (£21,962) was then determined and is shown in Figure 4. The Greggs-Pret lines are also displayed for comparison.

Qualitatively, the three lines follow similar paths. This could be because most wealth is concentrated in London, as are most of the Pret stores.

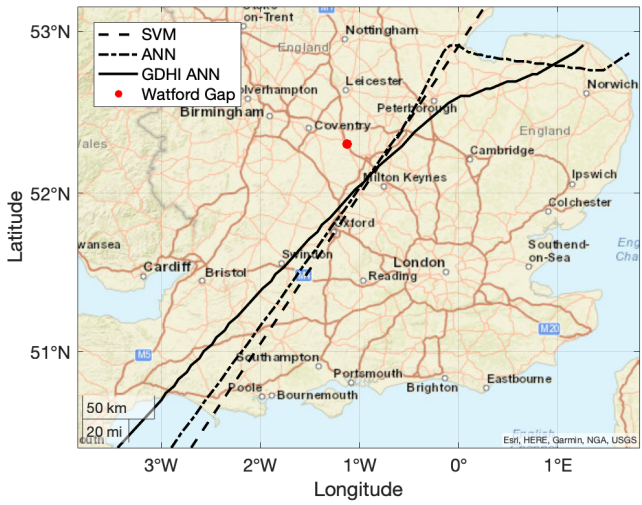

**Figure 4. The North-South dividing line as determined by the Artificial Neural Network regression analysis of GDHI data from 43 regions in England, compared with the predictions based on Greggs and Pret locations.**

### 3.3. Morrisons-Waitrose comparison

The Greggs-Pret geographical distribution is a flawed metric for a number of reasons. Firstly, Pret is a very London-centric brand meaning that it is possibly indicative of 'London-ness' rather than 'Southern-ness'. Additionally, neither of these brands appear often in the Southwest of England. Cornwall, the most south-westerly county in England, is famous for its Cornish pasties. Thus in this region Greggs does not have a monopoly in the baked goods sector* [11]. Therefore, the distributions of Morrisons and Waitrose supermarkets were instead considered. Neither brand is associated with one particular city and they permeate all corners of England; however, Morrisons is associated with 'Northern-ness' and Waitrose with 'Southern-ness'.

As before, 80% of the data were used for training the models and 20% were kept for testing. The trained SVM had an average accuracy of 64%, meaning that 64% of test

*Until 2019 no Greggs stores existed in Cornwall

data were correctly allocated into a Morrisons (North) and Waitrose (South) category. This lower accuracy makes sense, since stores for the two brands have larger overlaps geographically than Greggs and Pret. The ANN performed with 79% accuracy. The North-South dividing lines as derived by these models is shown in Figure 5. The line derived using the Artificial Neural Network falls within 2 miles of the Watford Gap.

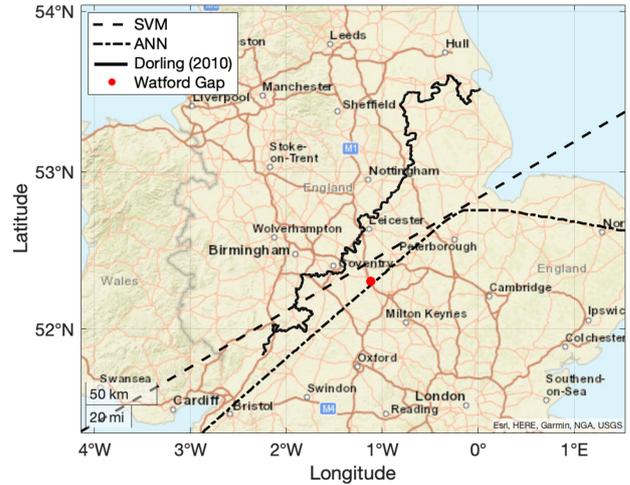

**Figure 5. The North-South dividing lines as determined by the distributions of Morrisons and Waitrose supermarkets. The line predicted by Dorling [12] is also plotted.**

## 4. Discussion

This light-hearted study is designed to highlight real and important cultural and socio-economic differences between the North and the South of England, and demonstrate that these are in some ways reflected by consumer habits. We attempted to align our Greggs-Pret lines with Gross Domestic Household Income with some success. However, this is more likely a reflection of the London-centricity of both England's wealth and its Pret-a-Manger shops. Additionally, this Greggs-Pret comparison determined that Cornwall – even including the most southern point in England – is actually Northern. For these reasons, the Greggs-Pret index appears to have significant limitations. As a result, we broadened the analyses to include the distribution of Morrisons and Waitrose stores and extracted alternative dividing lines.

In Figure 5, the results for Morrisons and Waitrose are compared with a line published by Dorling in 2010 [12]. Dorling performed a geographical analysis of England considering factors including life expectancy, education, unemployment, poverty levels, and house prices. Their result is a detailed line dissecting England on a political constituency-by-constituency basis. Our Morrisons-Waitrose lines have some overlap with the carefully considered line of Dorling. Since the work of Dorling in 2010, an explosion in the use of Machine Learning and AI has been witnessed across a great number of areas of research. We assert that an up-to-date re-analysis of data in the categories used by Dorling with modern methods could yield meaningful results.

## 5. Acknowledgements

R.S. thanks L. for their assistance in acquiring the geographical data and V.A.S.D. for proofreading.